\documentclass[10pt,aps,prl,twocolumn,amsmath,showpacs,amssymb,floatfix,superscriptaddress,longbibliography]{revtex4-1}
\usepackage{epsfig}
\usepackage{amsmath}
\usepackage{amssymb}
\usepackage{bm}
\usepackage{natbib}
\usepackage{graphicx}
\usepackage{color}
\usepackage{graphicx,rotating}
\usepackage{upgreek}
\usepackage[normalem]{ulem}

\begin{document}
\author{Thomas~Ding}\email{thomas.ding@mpi-hd.mpg.de}
\author{Marc~Rebholz}
\author{Lennart~Aufleger}
\author{Maximilian~Hartmann}
\author{Kristina~Meyer}
\author{Veit~Stoo{\ss}}
\author{Alexander~Magunia}
\author{David~Wachs}
\author{Paul~Birk}
\author{Yonghao Mi}
\author{Gergana Dimitrova Borisova}
\author{Carina da Costa Castanheira}
\author{Patrick Rupprecht}
\affiliation{Max-Planck-Institut f\"{u}r Kernphysik, Saupfercheckweg 1, 69117 Heidelberg, Germany}
\author{Zhi-Heng~Loh}
\affiliation{Division of Chemistry and Biological Chemistry, School of Physical and Mathematical Sciences, Nanyang Technological University, Singapore 637371, Singapore}
\author{Andrew~R.~Attar}
\affiliation{Department of Chemistry, University of California, Berkeley, California 94720, USA}
\author{Thomas~Gaumnitz}
\affiliation{Laboratorium f\"{u}r Physikalische Chemie, Eidgen\"{o}ssische Technische Hochschule Z\"{u}rich, Vladimir-Prelog-Weg 2, 8093 Z\"{u}rich, Switzerland}
\author{Sebastian~Roling}
\author{Marco~Butz}
\author{Helmut~Zacharias}
\affiliation{Physikalisches Institut, Westf\"{a}lische Wilhelms-Universit\"{a}t M\"{u}nster, Wilhelm-Klemm-Stra{\ss}e 10, 48149 M\"{u}nster, Germany}
\author{Stefan~D\"{u}sterer}
\author{Rolf~Treusch}
\affiliation{Deutsches Elektronen-Synchrotron DESY, Notkestra{\ss}e 85, 22607 Hamburg, Germany}
\author{Stefano~M.~Cavaletto}
\author{Christian~Ott}\email{christian.ott@mpi-hd.mpg.de}
\author{Thomas~Pfeifer}\email{thomas.pfeifer@mpi-hd.mpg.de}
\affiliation{Max-Planck-Institut f\"{u}r Kernphysik, Saupfercheckweg 1, 69117 Heidelberg, Germany}

\title{Nonlinear coherence effects in transient-absorption ion spectroscopy with stochastic extreme-ultraviolet free-electron laser pulses}

\begin{abstract}
We demonstrate time-resolved nonlinear extreme-ultraviolet absorption spectroscopy on multiply charged ions, here applied to the doubly charged neon ion, driven by a phase-locked sequence of two intense free-electron laser pulses. Absorption signatures of resonance lines due to 2$p$--3$d$ bound--bound transitions between the spin-orbit multiplets $^3$P$_{0,1,2}$ and $^3$D$_{1,2,3}$ of the transiently produced doubly charged Ne$^{2+}$ ion are revealed, with time-dependent spectral changes over a time-delay range of $(2.4\pm0.3)\,\text{fs}$. Furthermore, we observe 10-meV-scale spectral shifts of these resonances owing to the AC Stark effect. We use a time-dependent quantum model to explain the observations by an enhanced coupling of the ionic quantum states with the partially coherent free-electron-laser radiation when the phase-locked pump and probe pulses precisely overlap in time.

\pacs{.}
\end{abstract}
\maketitle
\begin{figure}[t]
\includegraphics{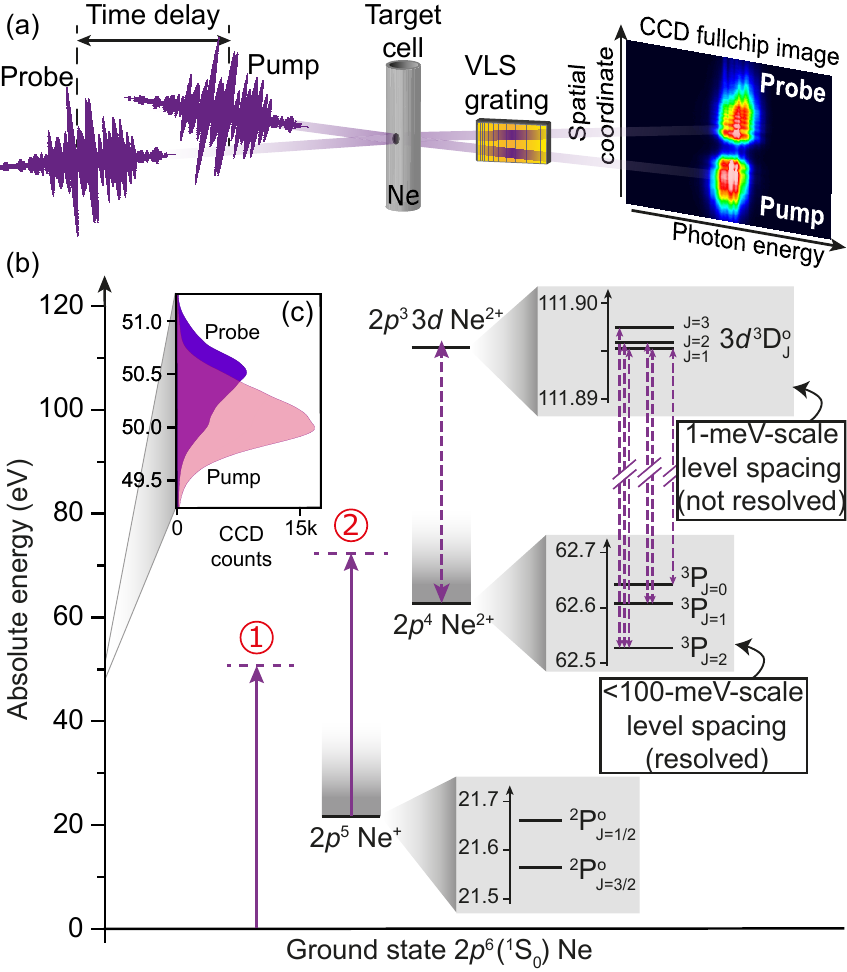}
\caption{(a) Experimental pump-probe geometry. (b) Energy level diagram and scheme of sequential single-photon absorption/ionization steps of neon at a photon energy of 50 eV (violet arrows), indicated by the red numbers (time-ordering from left to right). Axis of ordinate represents the absolute energy absorbed by the system with respect to the neon ground state. Ionization edges are shaded in gray. The levels in each charge state are denoted by their spin-orbit term symbol $^{2S+1}L_J$ with spin angular momentum $S$, orbital angular momentum $L$ and total angular momentum $J$. Energy levels drawn according to~\cite{NIST_Ne2016}. (c) Averaged FEL-photon spectra of pump and probe pulses.}
\label{fig_Ne_scheme}
\end{figure}
\begin{figure}[t]
\includegraphics{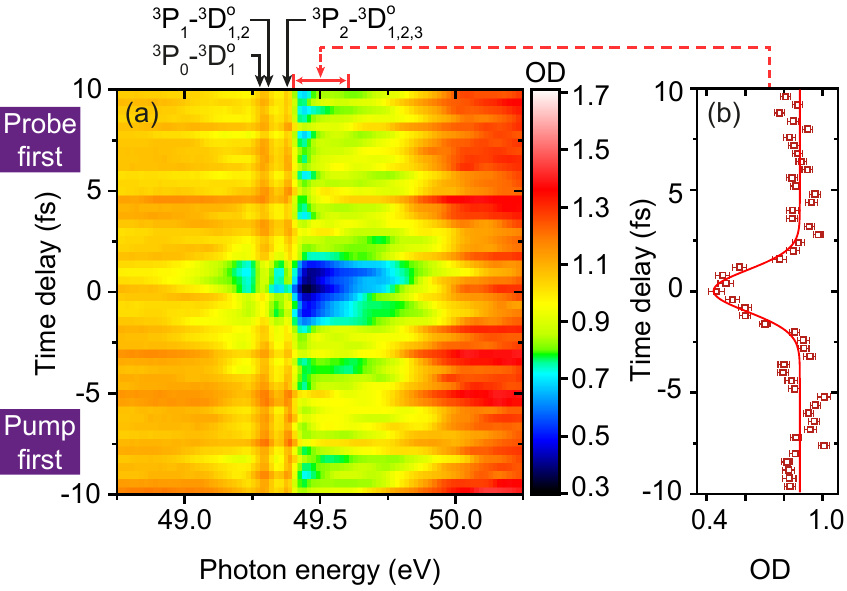}
\caption{(a) XUV-pump/XUV-probe transient absorption spectroscopy of neon measured with the probe spectrum. (b) Averaged lineout along the time-delay axis for the spectral range between 49.4 and 49.6 eV photon energy (square dots, error bars specify the standard deviation from the mean over 200 shots per delay setting) and Gaussian fit (red solid line).}
\label{fig_td-scan}
\end{figure}
\begin{figure}[t]
\includegraphics{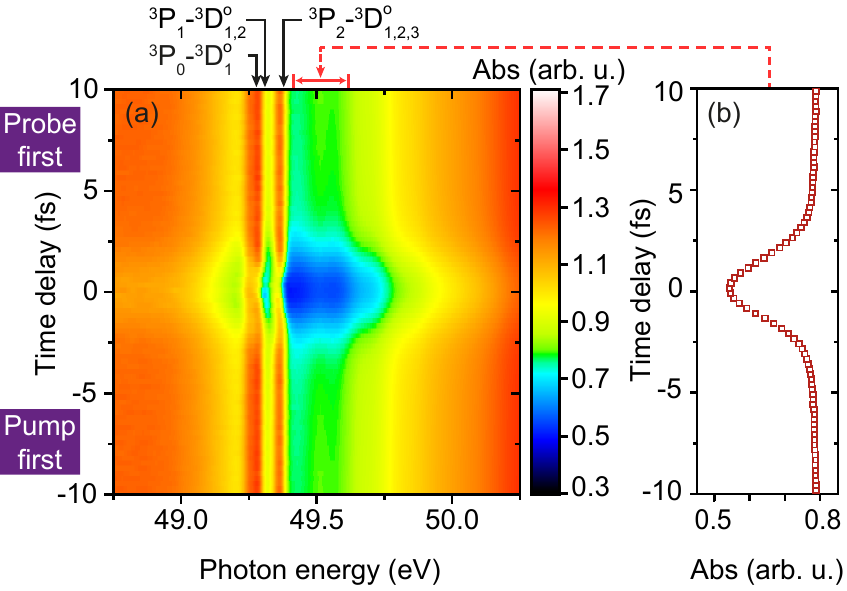}
\caption{(a) Numerical results obtained from a few-level coupling model of the experimentally observed $2p\,^3$P$_{0,1,2}-3d\,^3$D$_{1,2,3}$ transitions of the Ne$^{2+}$ ion. The absorption spectra are convolved with a Gaussian 30-meV FWHM spectrometer resolution of the experiment. Nonresonant absorbance changes due to plasma-diffraction effects are accounted for as described in the Supplemental Material (Section II). (b) Lineout along the time-delay axis for the spectral range between 49.4 and 49.6 eV photon energy. }
\label{fig_coherence_spike}
\end{figure}
\begin{figure}[t]
\includegraphics{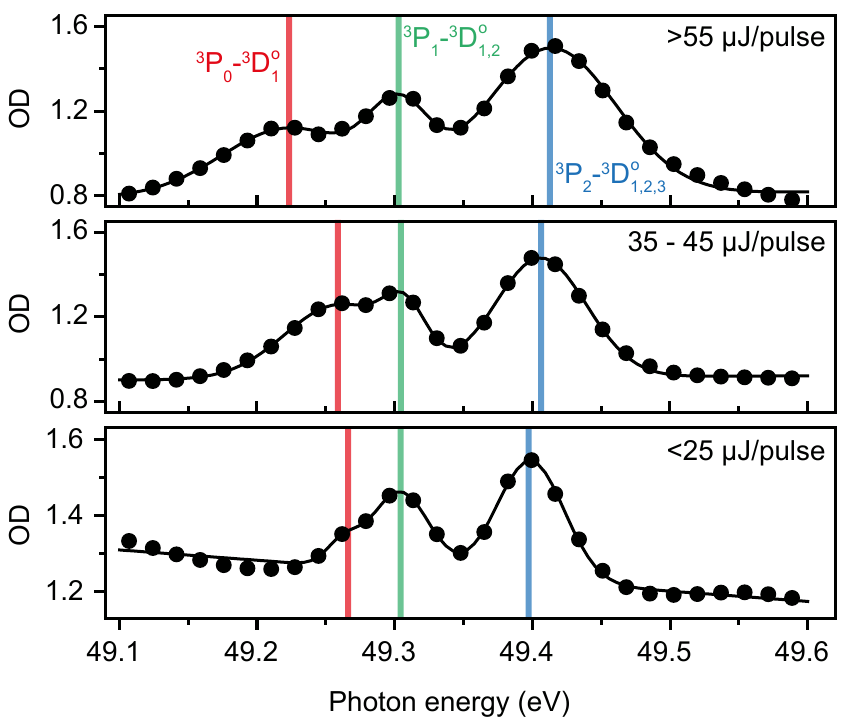}
\caption{Static XUV absorption spectrum of the $2p$--$3d$ $^3\text{P}$--$^3\text{D}$ transitions in Ne$^{2+}$ measured for three different FEL-pulse energy regimes. Dots are measured optical density (OD) values and black solid lines are fits with a triple-Gaussian model function. The colored vertical lines mark the fitted spectral positions of the multiplets (red $^3\text{P}_0$--$^3\text{D}_{1}$, green $^3\text{P}_1$--$^3\text{D}_{1,2}$, blue $^3\text{P}_2$--$^3\text{D}_{1,2,3}$).}
\label{fig_stark-shift}
\end{figure}

In interaction with matter the oscillating electric field of a laser not only induces transitions between bound electronic states but also affects the states and transitions themselves. It splits~\cite{Autler1955}, shifts~\cite{Fedorov1998,Delone1999} and modifies the width~\cite{Citron1977,Allen2012} and the shape~\cite{Ott2013,Kaldun2014,Ott2014} of spectral transition lines depending on the amount of detuning out of resonance with the laser frequency and the field strength. Only for sufficiently high field strengths, at which \emph{more} than one photon can interact with the quantum system on its intrinsic time and energy scale, these phenomena are accessible. Modern ultrafast lasers are effective driver and control tools for nonlinear effects at visible frequencies and have become the ``working horses'' for nonlinear coherent spectroscopies~\cite{MUKAMEL1995}, in time domain and frequency domain, including the quantum control of bound--bound electronic transitions (see, e.g.,~\cite{Brif2010} and references therein).\\ 
\noindent\hspace*{4mm}%
Since the advent of short-wavelength free-electron lasers (FELs)~\cite{Ackermann2007,McNeil2010} the field of nonlinear spectroscopy is being extended into the extreme-ultraviolet (XUV) and x-ray spectral ranges~\cite{Chapman2007,Sorokin2007,YOUNG2010,Kanter2011,Doumy2011,Rudek2012,Rohringer2012b,Weninger2013,Fuchs2015,Rudek2018,Young2018,Foglia2018}. One advantage of employing x-rays is the ability to access bound--bound electronic transitions associated with the spatially localized inner-electronic shell and the potential to probe site-specific spectroscopic information of a sample. Since experimental studies on the impact of XUV/x-ray nonlinear effects on inner-shell-excited resonances are often hampered by the extremely short Auger decay times, yet, such research is rare. Nonetheless, first x-ray nonlinear line-shape modifications of inner-shell transitions have been studied experimentally~\cite{Kanter2011} by employing Auger-electron spectroscopy. By contrast, we here address the valence electrons of the doubly charged neon ion, Ne$^{2+}$, and manipulate---in the absence of any competing ultrafast decay channel---the ground state to excited state transitions between spin-orbit multiplets with intense XUV-FEL radiation. Being sensitive to the atomic/ionic dipole response and associated spectral line-shape modifications, our work demonstrates a direct view on XUV nonlinear effects occurring in the strongly driven system involving bound--bound transitions.\\ 
\noindent\hspace*{4mm}%
In this Letter, we observe XUV-nonlinear physics near resonances in atomic ions. This is achieved by employing a time-resolved all-XUV nonlinear absorption spectroscopy method, based on the intense stochastic FEL pulses provided by the SASE (self-amplified spontaneous emission) Free-Electron Laser in Hamburg (FLASH). As a main result, we access the excited-state coherent response of the Ne$^{2+}$ ions with spin-orbit state-specific spectroscopic resolution and identify the 10-meV-scale resolved discrete ionic multiplet transitions. Taking advantage of the partial temporal coherence of the stochastic light fields, this approach allows one to capture transient effects of nonlinear coupling in the XUV spectra on a few-femtosecond timescale, well below the average FEL pulse duration, but within the eV-scale reciprocal (Fourier-inverse) average spectral bandwidth. Furthermore, we observe and quantify XUV-intensity-dependent spectral shifts of ionic resonances. The presented all-XUV-optical transient absorption experiment with combined state-resolved spectroscopic and time-resolved dynamic access to the nonlinear absorption response (at least third order) can be regarded as a precursor study to nonlinear multidimensional spectroscopy in the XUV/x-ray spectral domain~\cite{SCHWEIGERT2007,Bennett2016} with currently available SASE FEL technology.\\ 
\noindent\hspace*{4mm}%
In the experiment, we geometrically split the FEL beam into two approximately equal parts with average intensities in the mid $10^{13}\,\text{W/cm}^2$ region using the split-and-delay unit at beamline BL2~\cite{Wostmann2013}. Due to the beam's spatial coherence properties, a phase-locked linearly polarized pulse pair is produced on a shot-to-shot basis, which has been demonstrated and employed in several previous experimental campaigns~\cite{JIANG2010A,Jiang2010b,Moshammer2011,Roling2011,Meyer2012,Usenko2017}. Both pulses, denoted by pump and probe, respectively, are focused with a spot size of ($20\pm5$)$\,\upmu\text{m}$~\cite{Tiedtke2009} into a neon-filled gas cell ($\sim50\,\text{mbar}$ backing pressure) with $2\,\text{mm}$ interaction length. Further downstream, in the optical far field, the transmitted pump and probe pulses are simultaneously detected. They are separated via an offset in space, and spectrally resolved via a flat-field variable-line-spacing (VLS) grating in combination with a CCD camera, obtaining a resolving power $E/\Delta E\approx 1500$. Figure~\ref{fig_Ne_scheme}(a) provides a schematic illustration of the experimental setup. The FEL was operated in single-bunch mode at a $10\,\text{Hz}$ repetition rate and centered at 50.6~eV photon energy with $\sim0.8\,\text{eV}$ full width at half maximum (FWHM) spectral bandwidth of the average spectrum. The FEL pulse energy was measured shot by shot by the parasitic gas monitor detector (GMD)~\cite{Tiedtke2008} upstream of the experiment. This FEL operation mode allows for the (post-)analysis and sorting of the individually taken photon spectra with respect to the pulse energies. Averaging over all data points, the mean pulse energy was $47.5\,\upmu\text{J}$ with 28\% standard deviation due to the statistical shot-to-shot fluctuations. The average FEL temporal pulse duration was estimated to $\sim100\,\text{fs}$ based on the measurement of the electron bunch duration~\cite{Roehrs2009,Duesterer2014a}.\\
\noindent\hspace*{4mm}%
The physical mechanism of our experiment is illustrated in Fig.~\ref{fig_Ne_scheme}(b). The given FEL photon energy is sufficient to overcome the ionization thresholds~\cite{NIST_Ne2016} of the Ne atom at $21.6\,\text{eV}$, and its singly charged ion Ne$^{+}$ at $41.0\,\text{eV}$, respectively, via the sequential absorption of two XUV photons. The transiently created doubly charged Ne$^{2+}$ ions represent the actual target of this experiment, which are (near-)resonantly excited and identified via the 2$p$--3$d$ electronic bound--bound transitions between the spin-orbit multiplets $^3$P$_{0,1,2}$ and $^3$D$_{1,2,3}$. With the relatively large cross section for the sequential single-photon ionization processes ($\text{Ne}\to \text{Ne}^+$ with $\sigma_{01}=9.32$~Mb~\cite{Cavaletto2017} and $\text{Ne}^+\to \text{Ne}^{2+}$ with $\sigma_{12}=9.36$~Mb~\cite{Cavaletto2017} at 50 eV) we estimate a fractional Ne$^{2+}$ ion abundance of $\sim$70\%, based on coupled linear rate equations~\cite{Lambropoulos1987}. For this estimate, we assumed Gaussian pulses with 100 fs FWHM temporal duration and a peak intensity of $2\times10^{13}\,\text{W/cm}^2$ to model the average FEL pulse shape, relating to an incoming fluence of approximately $2\,\text{J/cm}^2$. One goal of this work is to investigate the coherent nonlinear few-femtosecond dynamics around the identified resonant transitions in Ne$^{2+}$ when both pump and probe pulses interfere and precisely overlap in time. Here it should be noted that the relevant ionic transitions are relatively long lived (on the timescale of the FEL pulse duration), hence their intrinsic decay dynamics are stationary during the 20-fs observation window. This enables an undisturbed access to the coherence effects of overlapping pump and probe pulses in the spectral vicinity of the resonances. We also identify enhanced plasma diffraction of XUV-optical pump photons into the direction of the detected probe beam, which is due to a nonlinearly enhanced Ne$^{2+}$ abundance in temporal overlap of pump and probe pulses, as another contribution to affect the measured probe absorption spectrum.\\
\noindent\hspace*{4mm}%
The measured absorbance $A$ is given in terms of the optical density (OD) following Beer--Lambert's law,
\begin{equation}\label{Transmission_function}   
A(\tau,\omega)=-\text{log}_{10}\bigg[\frac{I(\tau,\omega)}{I_0(\omega)}\bigg],
\end{equation} 
where $I(\tau,\omega)$ is the transmitted photon flux through the gas target, depending on both time delay and photon energy, and $I_0(\omega)$ is the incoming photon flux without the target, which was measured under the same experimental conditions. For weak-field optical transmission through the target, the quantity in Eq.~\ref{Transmission_function} is directly proportional to the single-photon absorption cross section~\cite{Gaarde2011a,Santra2011}. In order to minimize spectral irregularities due to the intrinsic FEL shot-by-shot fluctuations, the signal $I(\tau,\omega)$ was averaged over 200 consecutive single spectra at each time-delay-setting $\tau$, while the delay-independent incoming signal $I_0(\omega)$ was determined from the average over 3,300 FEL pulses. Note that spectral interferences according to the time delay of the two pulses are visible in the single-shot spectra but washed out in the averaged representation of $I(\tau,\omega)$ due to the 0.28~fs~\cite{Wostmann2013} shot-to-shot jitter of the interferometer. For details, see the Supplemental Material (section IV).\\
\noindent\hspace*{4mm}%  
Figure~\ref{fig_td-scan}(a) shows the time-delay-resolved absorbance of the probe pulse as it was scanned over the temporal pulse overlap (i.e., the delay setting $\tau=0\,\text{fs}$) with an incremental step size of $\Delta\tau=0.4\,\text{fs}$. Embedded in the continuum absorption background, dominant resonant signatures are observed at around 49.25~eV, 49.29~eV and 49.37~eV photon energy which can be identified as 2$p$--3$d$ bound--bound transitions between the spin-orbit multiplets $^3$P$_{0,1,2}$ and $^3$D$_{1,2,3}$ of Ne$^{2+}$, as schematically indicated by the dashed arrows in Fig.~\ref{fig_Ne_scheme}(b). Depending on $\tau$, the probe spectrum exhibits spectral changes including a decrease in the absorbance which is spectro-temporally localized near $49.5\,\text{eV}$ and when pump and probe pulses overlap perfectly in time around $\tau=0\,\text{fs}$. The measured temporal width of this transient bleach is $(2.4\pm0.3)\,\text{fs}$ (FWHM), as obtained from a Gaussian fit along the time-delay trace, cf. Fig.~\ref{fig_td-scan}(b). Time-delay-dependent changes of the continuum absorption background are not observed within the 20-fs observation window of the data shown in Fig.~\ref{fig_td-scan}. Such behavior can be observed over a longer time-delay range ($>50~\textrm{fs}$), as a step-like bleaching of the optical density, indicative of a reduction of the neutral species when the pump pulse fully precedes the probe pulse. See Supplemental Material (section V) for details. It should be noted that the observed spectral lines occur slightly out of resonance with the blue-detuned $(50.6\pm0.4)$-eV FEL photons of the probe beam and are thus measured only within the spectral wings.\\ 
\noindent\hspace*{4mm}%
In the following, we aim to develop a first understanding of the transient changes in the vicinity of the 2$p$--3$d$ resonances. For this purpose, we consider a few-level quantum model based on the approach previously described in the Supplemental Material of Ref.~\cite{Ott2014}, which we now employ with stochastic XUV fields to nonlinearly drive the coupled $^3\text{P}$--$^3\text{D}$ $2p$--$3d$ multiplet transitions of Ne$^{2+}$. Pump and probe electric fields are treated in the framework of the partial-coherence method~\cite{Pfeifer2010} as two identical copies (perfect phase lock) with identical random-phase characteristics and delayed by $\tau$ with respect to each other. The starting point of this few-level model simulation is the $^3$P$_{0,1,2}$ ionic ground state which we assume to be initially incoherently populated with equal probability across the spin-orbit substates owing to the statistical nature of the FEL pulses. This assumption also suppresses the formation of any spin-orbit wave-packet dynamics of the ionic ensemble in agreement with the experimental observation. Although we neglect any effect of the slow, 100-fs-timescale $\tau$-dependent buildup of the Ne$^{2+}$-ion itself, we combine the few-level model with rate equations~\cite{Meyer2012} to account for the few-fs-timescale nonlinear coherence enhancement effect of the peaking Ne$^{2+}$-population yield at around $\tau=0\,\text{fs}$. This leads to an increased diffraction of pump photons in the direction of the probe beam and effectively reduces the absorbance of the probe. The effect is most pronounced in the 49.5 eV energy region where a shift to slightly lower photon energy of the pump is most significant (see Fig.~\ref{fig_Ne_scheme}c). A detailed description of the numerical approach can be found in the Supplemental Material (Sections I and II).\\   
\noindent\hspace*{4mm}%
The computational results are shown in Fig.~\ref{fig_coherence_spike}. As in the experiment (cf. Fig.~\ref{fig_td-scan}), the simulated pump--probe absorbance trace exhibits a spectral bleach in the vicinity of the resonances near 49.5 eV photon energy at around $\tau=0\,\text{fs}$, when pump and probe pulses precisely overlap in time and interact coherently. This is due to (i) additional (enhanced) nonlinear coupling between the near-resonantly driven transitions and (ii) enhanced plasma diffraction of the pump pulse as a consequence of the constructive interference of the ``spiky'' temporal sub-structure of the replica pulses (cf. Fig. S1 of the Supplemental Material). The qualitative agreement between experiment and simulation proves direct access to nonlinear coherence effects around these ionic transitions with this all-XUV-optical experimental method.\\
\noindent\hspace*{4mm}% 
The observed trend of an enhanced coherent coupling in the pump--probe absorbance trace at zero time delay is similar to those discussed in early time-dependent transient-grating experiments with optical lasers~\cite{VARDENY1981, EICHLER1984}, where the transient peaking in the measurement signal is explained by the nonlinear (wave-mixing) nature of the light--matter interaction. When combined with temporally incoherent fields, the accessible time resolution is essentially determined by the fields' coherence time~\cite{Morita1984,Tomita1986}. The coherence time is quantified by the temporal width of the signal enhancement, provided that the replica pulses share identical spectro-temporal properties. The here observed $(2.4\pm0.3)\,\text{fs}$ timescale correlates with the measured $\sim0.8$~eV FWHM spectral bandwidth of the average spectrum according to the time--bandwidth product of Fourier-limited Gaussian pulses. It thus serves as an order-of-magnitude estimate for the coherence time.\\
\noindent\hspace*{4mm}% 
Finally, to systematically study the role of the nonlinear mechanisms that contribute to the changes of the electronic structure of the Ne$^{2+}$ ion, we now perform \emph{static} XUV absorption spectroscopy and vary the FEL-pulse intensity. In this case, the fields are temporally separated in the target ($\tau>100$ fs, probe first) and the transmitted spectra of the succeeding (pump) pulse are analyzed. While the preceding probe pulse already produces a high abundance of the Ne$^{2+}$ ions, the succeeding pump pulse acts as a moderately strong dressing field of the ionic states and its optical absorption response is directly measured. The role of pump and probe are thus interchanged in this static setting, where the slightly weaker ``probe'' pulse is still strong enough to create a substantial amount of Ne$^{2+}$ ions, while the temporally separated and stronger ``pump'' pulse is used to induce nonlinear effects. In Fig.~\ref{fig_stark-shift} we show the measured XUV absorption spectra of the previously discussed ionic resonance features near 49.3 and 49.4 eV photon energy, now resolved for different FEL pulse energy regimes as selected by the GMD shot-by-shot measurement. Compared to the ``natural'' neon ionic spectrum, as obtained from high-power discharge-source measurements~\cite{Livingston1997}, the results presented here reveal considerable light-induced spectral modifications including clear intensity-dependent spectral shifts with a maximum relative resonance shift of about 50 meV observed for the $^3\text{P}_0$--$^3\text{D}_{1}$ transition when subjected to an average FEL pulse energy of $>55\upmu\text{J}$ (marked by the red vertical lines in Fig.~\ref{fig_stark-shift}). This observation suggests nonlinear coupling dynamics of ionic resonances with strong XUV electric fields, pertaining to the dynamic Stark shift. It is interesting to see that the three resonances of the spin-orbit multiplet actually shift in different directions, effectively increasing their energy splitting, while the FEL photon energy centered at $\sim50.6$~eV is blue-detuned to all resonances. Besides the repulsion of all resonances, we observe a shift in the mean position of the multiplet structure toward a lower transition energy, which is evidenced by the stronger redshift of the lowest resonance (red) with a comparatively weaker blueshift of the highest resonance (blue), while the central line (green) remains unchanged. Furthermore, we observe the lowest transition (red line) to nonlinearly increase the strength of its shift. We expect the nearby $^3\textrm{S}_1$ state at 52~eV, for which the FEL photon energy appears red-detuned, to play a major role in this seemingly complicated AC-stark-shifted multiplet structure. A numerical analysis of the observed shifts, including the coupling to the $^3\textrm{S}_1$ state, is presented in the Supplemental Material (Section III). This first experimental observation of such non-trivial light-induced level shifts clearly highlights the prospect of the here presented method of strong-field XUV absorption spectroscopy even in the static case, with direct access to distinct relativistic spin-orbit transitions driven by strong electric fields.
\\
\noindent\hspace*{4mm}% 
In conclusion, we have presented first experimental measurements of XUV-induced nonlinear coherence effects near excited-state ionic resonances with overlapping FEL pump and probe pulses in transient absorption geometry. The measured time scale of this transient effect is $(2.4\pm0.3)\,\text{fs}$. The sensitivity to transient changes combines high spectral (limited by the grating spectrometer) with high temporal (limited by the FEL coherence bandwidth) resolution. The observed level shifts in strong XUV electric fields demonstrate the broad applicability of this nonlinear spectroscopy method with no fundamental limitation on spectral resolution, despite the much larger FEL bandwidth. The experiment thus represents a significant step towards the implementation of coherent multi-dimensional spectroscopy of XUV-excitation and decay dynamics in atoms and molecules with broadband intense SASE FELs.

\begin{acknowledgments}
We thank Z.~Harman for helpful discussions and for providing values of the dipole-moment matrix elements.
\end{acknowledgments}

\bibliography{Ding_et_al_manuscript_to_PRL}
\end{document}